# Quasi-one-dimensional electron gas bound to a helium-coated nanotube

Michael Liebrecht[1,2,3,4], Adrian Del Maestro[5] and Milton W. Cole[1]

[1]Department of Physics, Penn State University, 104 Davey Laboratory, University Park, PA 16802
[2]Johann Radon Institute for Computational and Applied Mathematics, Austrian Academy of Sciences, Altenberger Straße 69, A-4040, Linz, Austria
[3]MathConsult GmbH, Altenberger Straße 69, A-4040, Linz, Austria
[4]Institute for Theoretical Physics, Johannes Kepler University, Altenberger Straße 69, A-4040, Linz, Austria
[5]Department of Physics, University of Vermont, Burlington, VT 05401

Abstract

A much-studied system is the quasi-2D electron gas in image-potential bound states at the surface of helium and hydrogen. In this paper, we report on an analogous quasi-1D system: electrons bound by image-like polarization forces to the surface of a helium-coated carbon nanotube. The potential is computed from an electron-helium pseudopotential, plus a dynamic image term evaluated from a semi-classical model of the nanotube's response function. Predictions are made for the bound states and potential many-body properties of this novel electron gas for a specific choice of tube radius and film thickness.

## A. Introduction

An ongoing focus of research in many-body and low temperature physics is the behavior of systems characterized by reduced dimensionality. Exploration of this subject has motivated vast research concerning monolayer films on flat surfaces as well as on the properties of electrons within graphene, both of which are quasi-two-dimensional (2D) systems [1,2]. Here the prefix "quasi" refers to the fact that only two spatial coordinates of the particles extend to infinity, so that phase transitions can exhibit finite T, infinitely long-range order in 2D. More recently, attention has been drawn to the problem of quasi-1D systems (*i.e.,* just one divergent coordinate), exemplified by the physics of nanotubes, made of carbon or nanoporous materials, such as FSM-16, zeolites and MCM-41 [3]. Quasi-1D systems are of particular importance, as they are expected to manifest unique behavior that can be universally described via the Tomonaga-Luttinger liquid theory (See [4] for a recent review). Fundamental interest in low dimensional systems arises from the intriguing effects of dimensionality *per se* as well as their inherent tunability through the alteration of the pore size, film thickness or other geometrical or compositional parameters. A key aspect of each problem is the species of particle under nano-confinement (*e.g.* electron, atom or molecule), as the various physical phenomena involve inter-particle interactions in a non-trivial way.

One of the first kinds of quasi-2D electronic system to be explored was the electron bound to the surface of liquid helium (or solid hydrogen or neon) in a so-called *image state*. These electrons were predicted and observed to be quite weakly bound (~ 1 meV~ 10 K), with a mean electron-surface separation of approximately 10 nm, implying that they translate essentially freely parallel to the substrate, with scattering due to ripplons and gas phase atoms [1]. The electron is excluded from the liquid because of a short-range electron-atom repulsion but it is attracted to it by a long-range polarization (induced dipole) force. In the case of a very low-density electron gas (spacing large compared to the height above the film and T not extremely low), the gas is a quasi-2D, nearly free electron system. More interesting behavior occurs at higher density, where significant

many-body effects, such as Wigner crystallization, arise and enrich the exhibited 2D physics. [5,6]

This paper proposes a novel *quasi-1D* electronic system, related to the same principles that appear in the image state problem. Consider an infinitely long carbon nanotube, upon which has been deposited a thin helium, hydrogen or neon film (of order one or two layers). As in the image state, the electron is repelled by the film and therefore must spend its time (speaking classically) outside of the domain occupied by the nanotube. It is weakly bound to the tube-film complex by polarization forces analogous to those present in the image problem. As a result, it becomes bound in image-like states possessing the cylindrical symmetry associated with the nanotube. A schematic depiction of these states appears in Fig. 1, where the z axis is along the nanotube. Analogous to the 2D image state, an isolated electron in this geometry is nearly free to move in the z-direction. Here, too, the role of interactions with other electrons and excitations of the nanotube and film presents interesting behavior.

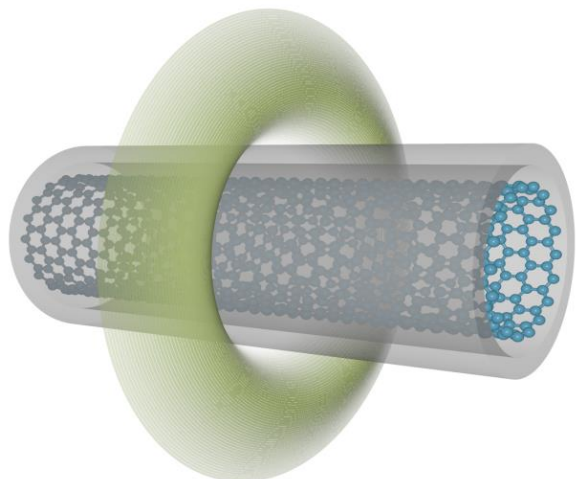

**Figure 1: A schematic picture of a single electron in its radial groundstate wavefunction (green torus) above a carbon nanotube covered by a single layer helium film (grey cylinder). The nanotube is aligned with the z-axis. The electron's localization along the z-direction is achieved by assembling a packet with a range of $k_z$ vectors.**

This paper presents the methodology and some initial results for these states. The potential and wave functions of the single electron problem are described, along with the nature of the electron-electron interaction, which is quite unusual. Comments about the novel physics arising in the many-body system are also included.

**B. Single electron problem**

Consider the system depicted in Fig. 1, where, due to symmetry, we use cylindrical coordinates $(\rho,\varphi,z)$ with basis vectors denoted by $\boldsymbol{e}_\rho$, $\boldsymbol{e}_\varphi$ and $\boldsymbol{e}_z$. We assume that the nanotube as well as the surrounding helium layers can each be described as homogeneous, infinitely thin cylindrical shells; corrections due to thickness and atomicity turn out to be small because of the large value of $\langle\rho\rangle$. The one-electron Schrodinger equation is

$$\left[-\frac{\hbar^2}{2m_e}\left(\frac{1}{\rho}\frac{\partial}{\partial\rho}\rho\frac{\partial}{\partial\rho}+\frac{1}{\rho^2}\frac{\partial^2}{\partial\varphi^2}+\frac{\partial^2}{\partial z^2}\right)+v_e(\rho)\right]\psi(\rho,\varphi,z)=E\,\psi(\rho,\varphi,z)\,. \quad (1)$$

Here $m_e$ is the electron mass, $v_e(\rho)$ is the potential caused by the coated nanotube and $\psi(\rho, \varphi, z)$ is the electron's wave function. Since $v_e(\rho)$ depends only on the radial coordinate, we write the solution of Eq. 1 as a product of radial, azimuthal and plane wave functions:

$$\psi(\rho, \varphi, z) = R_{n,\mu}(\rho) \frac{1}{\sqrt{2\pi}} e^{i\mu\varphi} e^{ik_z z} . \tag{2}$$

The resulting one-dimensional Schrodinger equation for given values of $\mu$ and $k_z$ (longitudinal wave vector) is

$$\left[ -\frac{\hbar^2}{2m_e} \frac{1}{\rho} \frac{\partial}{\partial \rho} \rho \frac{\partial}{\partial \rho} + \frac{\hbar^2 \mu^2}{2m_e \rho^2} + v_e(\rho) \right] R_{n,\mu}(\rho) = \varepsilon_{n,\mu} R_{n,\mu}(\rho) \tag{3}$$

$$E_{n,\mu}(k_z) = \varepsilon_{n,\mu} + \frac{\hbar^2 k_z^2}{2m_e} . \tag{4}$$

Thus the spectrum of states is a sum of radial and longitudinal energies. The potential $v_e(\rho)$ is taken as a sum of two contributions,

$$v_e(\rho) = v_{e-NT}(\rho) + v_{e-film}(\rho) , \tag{5}$$

where $v_{e-NT}(\rho)$ and $v_{e-film}(\rho)$ are the potentials due to the nanotube and its coating film, respectively. The omission of many-body corrections to this additive approximation is justified by the fact that screening by the film is small since its polarizability is small. The film's contribution to $v_e(\rho)$ is similarly derived by assuming that it is an integrated sum of individual e-atom induced-dipole interactions. For the nanotube's contribution, we also use this pairwise sum approximation together with effective, anisotropic carbon polarizabilities taken from flat graphene sheets [7] to take screening into account. The polarizability for fields perpendicular and parallel to the graphene surface are denoted by $\alpha_\perp = 0.87 Å^3$ and $\alpha_\parallel = 2.47 Å^3$, respectively. Thus the polarizability tensor of a carbon atom in a nanotube can be written as

$$\bar{\alpha}_{NT} = \alpha_\perp \boldsymbol{e}_\rho \otimes \boldsymbol{e}_\rho + \alpha_\parallel \boldsymbol{e}_\varphi \otimes \boldsymbol{e}_\varphi + \alpha_\parallel \boldsymbol{e}_z \otimes \boldsymbol{e}_z . \tag{6}$$

The electric field at $\vec{r}$ caused by an electron at position $(\rho,0,0)$ outside the nanotube is given by

$$\boldsymbol{E}(\vec{r}, \rho) = \boldsymbol{E}(r, \varphi, z, \rho) = -e \frac{(\rho - r\cos\varphi) \cdot \boldsymbol{e}_\rho + r\sin\varphi \cdot \boldsymbol{e}_\varphi + z \cdot \boldsymbol{e}_z}{(\rho^2 - 2\rho r \cos\varphi + r^2 + z^2)^{3/2}} , \tag{7}$$

where $e$ is the electron charge. Matter near the electron responds to its field by forming induced dipoles, which couple to the original field and cause an attractive interaction of the form

$$u(\vec{r}, \rho) \equiv -\frac{1}{2} \boldsymbol{E}(\vec{r}, \rho) \cdot \bar{\alpha}(\vec{r}) \cdot \boldsymbol{E}(\vec{r}, \rho) . \tag{8}$$

Here $\bar{\alpha}(\vec{r})$ is the polarizability tensor of the surrounding matter and the factor ½ accounts for the induced nature of the interaction. The total potential seen by the electron equals

$$v(\rho) = \int d^3r \, u(\vec{r}, \rho) \, n(\vec{r}) , \tag{9}$$

where $n(\vec{r})$ represents the particle density of the surrounding matter. Inserting the carbon density of a nanotube with radius $R_{NT}$, $n_{NT}(\rho, \varphi, z) = n_{s,C} \, \delta(\rho - R_{NT})$, as well as Eq. (6) into Eq. (9) yields

$$\begin{aligned} v_{e-NT}(\rho) = & -\frac{e^2 \pi n_{s,C}}{8 R_{NT} (\rho - R_{NT})^2 (\rho + R_{NT})} \\ & \times \Big[ \big( (3R_{NT}^2 + \rho^2) E(k_{NT}(\rho)) - (\rho - R_{NT})^2 K(k_{NT}(\rho)) \big) \alpha_\parallel \\ & + \big( (5R_{NT}^2 - \rho^2) E(k_{NT}(\rho)) + (\rho - R_{NT})^2 K(k_{NT}(\rho)) \big) \alpha_\perp \Big] \end{aligned} \tag{10}$$

with

$$k_{NT}(\rho) \equiv \frac{2\sqrt{R_{NT} \, \rho}}{R_{NT} + \rho} . \tag{11}$$

The functions $K(k)$ and $E(k)$ denote the complete elliptic integral of the first and second kind, respectively. To obtain the attractive potential of a $^4$He shell, we simply have to replace $R_{NT}$ with $R_{He}$, the radius of the helium shell and $n_{s,C}$ with $n_{s,He}$, the helium density, as well as substitute the isotropic helium polarizability $\alpha_{He} = 0.205 \text{Å}^3$ [8] for $\alpha_\perp$ and $\alpha_\parallel$. This results in

$$v^{(a)}_{e-shell}(\rho, R_{He}, n_{s,He}) = -\frac{e^2 \pi R_{He} n_{s,He} \alpha_{He}}{(\rho - R_{He})^2 (\rho + R_{He})} E\big( k_{He}(\rho, R_{He}) \big) \tag{12}$$

with

$$k_{He}(\rho, R_{He}) \equiv \frac{2\sqrt{R_{He} \, \rho}}{R_{He} + \rho} . \tag{13}$$

We now address the hard-core repulsion of the e-helium potential, which is ultimately responsible for the barrier excluding the electron from the fluid. This barrier may be represented by adding a model repulsion to Eq. (8):

$$u_{e-He}(\rho, \varphi, z) \equiv \frac{1}{2} \alpha_{He} \boldsymbol{E}(\rho, \varphi, z) \cdot \boldsymbol{E}(\rho, \varphi, z) \left[ \frac{a^4 \boldsymbol{E}(\rho, \varphi, z) \cdot \boldsymbol{E}(\rho, \varphi, z)}{2 e^2} - 1 \right] . \tag{14}$$

Here a is the equilibrium distance of the potential. If we choose $a = 1\text{Å}$, the resulting minimum of $|u_{e-He}|$ is about 0.7 eV, which is intuitively reasonable. In the present application, as in the flat surface case, this repulsive wall is necessary but the energy is not very sensitive to the functional form of the repulsion or the value of the parameter a. Applying Eq. (9) to the modified interaction specified in Eq. (14) yields the e-shell potential

$$v_{e-shell}(\rho, R_{He}, n_{s,He}) \equiv v^{(a)}_{e-shell}(\rho, R_{He}, n_{s,He}) + v^{(r)}_{e-shell}(\rho, R_{He}, n_{s,He}) , \tag{15}$$

which includes the previously discussed attractive part $v^{(a)}_{e,shell}(\rho, R_{He}, n_{s,He})$, from Eq. (12), and the repulsive part

$$v^{(r)}_{e-shell}(\rho, R_{He}, n_{s,He}) = \frac{a^4 e^2 \, \pi \, R_{He} n_{s,He} \alpha_{He}}{48(\rho - R_{NT})^6 (\rho + R_{NT})^5} \\ \times \left[ (23 R_{He}^4 + 82 R_{He}^2 \rho^2 + 23\rho^4) E(k_{He}(\rho, R_{He})) \\ - 8(\rho - R_{NT})^2 (\rho^2 + R_{He}^2) K(k_{He}(\rho, R_{He})) \right]. \tag{16}$$

The interaction between the electron and the full helium film consisting of N helium shells with radii $R_{He}^{(i)}$ and constant densities $n_{s,He}^{(i)}$ can then be constructed as sum of e-shell potentials:

$$v_{e-film}(\rho) = \sum_{i=1}^{N} v_{e-shell}\left(\rho, R_{He}^{(i)}, n_{s,He}^{(i)}\right) \tag{17}$$

Fig. 2 presents numerical results for the potential for the case of a single layer $^4$He film on a tube of radius 7.09Å. We assumed a separation distance between the nanotube and the He film of 2.90Å, $n_{s,C} = 0.38$Å$^{-2}$ and $n_{s,He} = 0.12$Å$^{-2}$. The calculations were performed on a logarithmic radial grid ranging from 5Å to $10^{10}$Å using the Numerov integrator [9-10] combined with a shooting method. The discretization is chosen to be equally spaced in logarithmic space and such that any interval $[\rho, 10\rho)$ contains 2500 grid points.

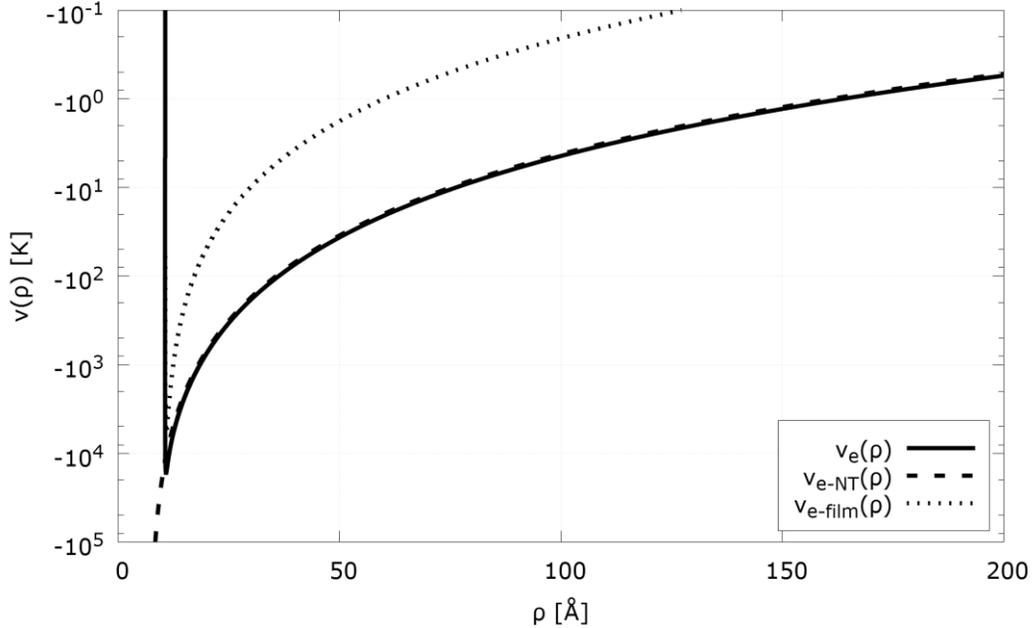

**Figure 2: Potential energy experienced by an electron as a function of radial distance in the case of a 1 layer film. The dashed curve (dotted curve) represents the nanotube's (film's) contribution to the total potential (full curve).**

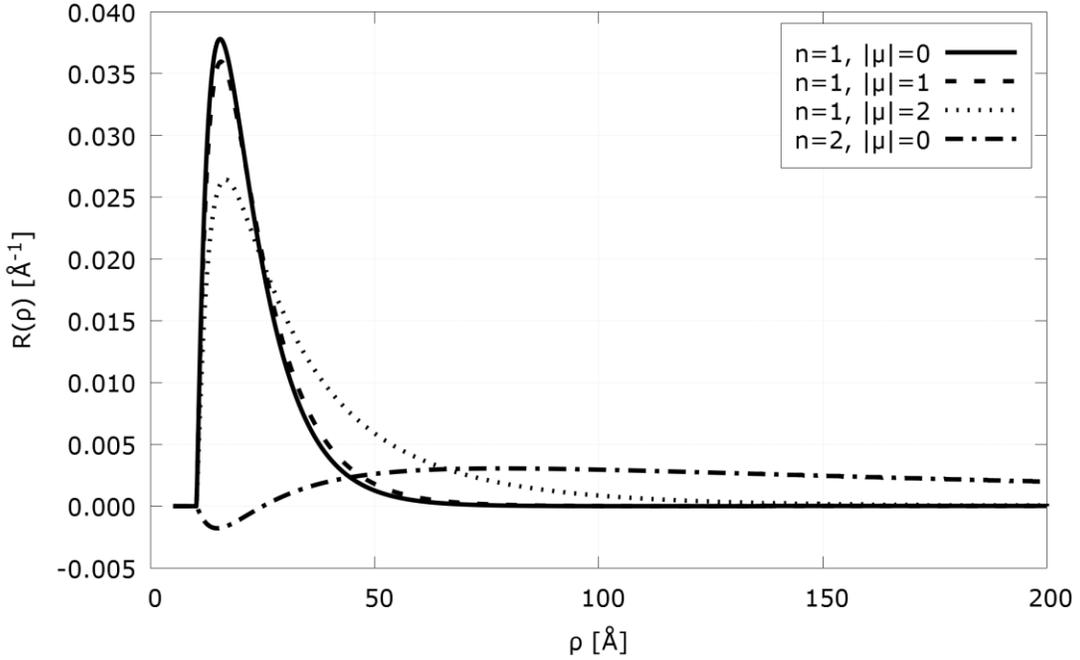

**Figure 3: Radial wavefunctions of the four lowest energy states with quantum numbers n=1, μ=0 (full curve), n=1, |μ|=1 (dashed curve), n=1, |μ|=2 (dotted curve) and n=2, |μ|=0 (dash-dotted curve). The corresponding eigenenergies are $\varepsilon_{1,0}$=-490.68K, $\varepsilon_{1,1}$=-360.33K, $\varepsilon_{1,2}$=-28.58K and $\varepsilon_{2,0}$=-0.41K. These six states are the only bound states of the potential shown in Figure 2.**

As seen in Fig. 3, there exist strongly bound states for this problem, where the "benchmark" reference is the case of an electron on the semi-infinite He liquid, with binding energy about 7 K. The much stronger binding in the present case is a consequence of the enhanced attraction provided by the nanotube, greatly exceeding that in the benchmark problem. The reason for the difference is that the polarizability of the C atoms in the fullerene is about a factor of 10 greater than that of a He atom. As a result, the electron is strongly attracted to the tube-film complex. In the case of a single layer He film, there are found 6 bound states, five of which arise from the lowest radial solution of Eq. (3), for azimuthal quantum numbers μ=0, ±1 and ±2, while one is the single excited state for μ=0. The energy differences between the former states can be estimated approximately by perturbation theory from the expectation value of the (cylindrical) centrifugal potential in Eq. 3: $\Delta_{\mu 0} \equiv \varepsilon_{\mu 0} - \varepsilon_{00} \sim [\hbar^2/(2m_e \rho^2)]\mu^2$. For μ=1, estimating $\langle 1/\rho^2 \rangle \sim (2\text{ nm})^{-2}$, this expression yields $\Delta_{10} \sim$ 110 K, not very different from the actual energy difference of 130 K; the calculation is suspect when the perturbation is so large. Results for other assumed values of the nanotube radius and film thickness will be published in a more complete publication.

### C. Tomonaga-Luttinger liquid theory

The existence of bound states above the surface of the helium film presents the intriguing possibility of engineering an extremely clean quasi-1D electron system. It is well known that in 1D, even at T = 0 K, there is no broken continuous symmetry, but instead the persistence of only quasi-long range order characterized by algebraic decay of correlation functions [11]. Regardless of the microscopic details, systems displaying these features are known as Tomonaga-Luttinger liquids (TLLs) [12-15] and can be described at low energies and long wavelengths by a single universal Hamiltonian (for spinless particles) describing the dynamics of two phase fields $\phi(z,t), \theta(z,t)$:

$$H_{TLL} = \frac{\hbar v}{2\pi} \int_0^L dz \left[\frac{1}{K}\left(\frac{\partial \phi}{\partial z}\right)^2 + K\left(\frac{\partial \theta}{\partial z}\right)^2\right] \quad (18)$$

where $\partial_z \theta$ is the canonically conjugate momentum to $\phi$. Excitations are phonon-like and propagate at velocity v while the Luttinger parameter $K$ is equal to unity for non-interacting electrons and takes values $K < 1$ for repulsive and $K > 1$ for attractive (repulsive) interactions. The particular values of v and $K$ are non-universal and depend on the microscopic details of the 1D system under consideration.

The amazing feature of $H_{TLL}$, which is a specialty of 1D quantum mechanics, is that interactions between electrons simply renormalize $K$ at low energies, with the effective theory describing linear quantum hydrodynamics. This form allows for the explicit calculation of many-body observables, including correlation functions, and there is great interest in testing its validity in real experimental systems of interacting particles. For electrons, TLL behavior has been observed in the tunneling and conductance of carbon nanotubes [16], via Coulomb drag in quantum wires [17] and in quantum Hall edge states [18-19]. However, such searches are often complicated by the fact that disorder or the presence of a periodic potential can "pin" a TLL, drastically altering its transport properties.

The system of electrons bound to helium proposed here would be mostly insensitive to these effects and it is natural to ask if it can exhibit the properties of a TLL. It is thus important to understand how "one-dimensional" the bound electrons are, and as shown in Figure 3, the ground state with angular quantum number $\mu = 0$ sits $\Delta_{10} \cong 130.3$ K below the first excited state ($\mu = 1$). At low temperatures, in the absence of interactions, the electrons can be assumed to be one-dimensional provided that $d_e k_F \ll 1$, where $d_e \approx 30$Å is radial diameter of the bound state wavefunction and $k_F = \pi n_{1D}/2$ is the Fermi wavevector with $n_{1D}$ the number of adsorbed electrons per unit length. This sets an upper bound on the density of $n_{1D} \approx 100$ μm$^{-1}$ which is on the order of densities achieved in pristine quantum wires [17]. Turning on Coulomb interactions between electrons, we require that higher angular momentum states remain unexcited (for z-localized electrons) and by setting $V_{e-e}(n_{1D}^{-1}) = e^2 n_{1D}/4\pi\varepsilon_0 = \Delta_{10}$ we find $n_{1D} \lesssim 10$ μm$^{-1}$ which is potentially achievable in adsorbed electron systems. Considering localized electron wavepackets, it is possible that interactions could be further suppressed by the unique geometry of the cylindrical bound states. The simple estimates presented here should be made more precise with a scattering calculation employing the single particle wavefunctions, which we postpone for future work.

The implications of such 1D behavior are exciting, and would be manifest in a specific heat that is linear in temperature, and exotic transport properties, including a non-linear conductance and possible spin-charge separation as measured by tunneling experiments [20].

### D. Comments and conclusions

The novel states proposed here are of special interest to the 1D transport and phase transition communities of scientists. The proposed states should exhibit high conductivity for the case of an electric field along the z axis, with mobility limited by e-e interactions, e-gas atom interactions and e-film ripplon interactions, as in the 2D case. Since the cylindrical equivalent of the Bohr radius is so high, we predict a large diamagnetic susceptibility, since that quantity is proportional to $<\rho^2>$. The principal experimental difficulties will arise from both the problem of maintaining charge and in creating sufficiently many coated nanotubes so as to permit experimental study, such as spectroscopy, which in principle is an ideal tool for testing our model. We look forward to hearing results from creative experimental exploration of this system.

## 5. Acknowledgments

We are grateful to S. Rotkin and S. Hernandez for several helpful conversations. M. Liebrecht gratefully acknowledge funding from the Austrian Science Fund FWF (project P21924 to Eckhard Krotscheck) and the Austrian Marshall Plan Foundation.